# Meson-meson interactions – from static to dynamic valence quarks[*]

H.R. Fiebig[a], H. Markum, A. Mihály, K. Rabitsch, W. Sakuler and C. Starkjohann[b]

[a]Physics Department, F.I.U. - University Park, Miami, Florida 33199, USA

[b]Institut für Kernphysik, Technische Universität Wien, A-1040 Vienna, Austria

A method for the extraction of an effective meson-meson potential from Green functions, which can be obtained from a lattice simulation, is presented. Simulations are carried out for compact QED and QCD in four dimensions using the quenched approximation and the hopping parameter expansion. In a further study, a heavy-light meson is considered employing a conjugate gradient algorithm for the light propagators. Due to the Pauli exclusion principle, the results for QED indicate the existence of a hard core, but for QCD there is strong attraction at small meson distances.

## 1. INTRODUCTION

QCD is believed to be the correct theory to describe the physics of strong interaction. Therefore it should be possible, in principle, to extract the potential between hadrons from this theory. Before we can proceed, we have to define what is meant by "potential". Throughout this paper we will use two different definitions:

1. The potential is the function $V(r)$ that has to be inserted into a Schrödinger equation so that quantum mechanics and quantum field theory result in the same predictions.

2. The potential at finite temperature is approximated by the thermodynamic free energy $F(r)$.

For the first definition we will carry out a simulation at zero temperature and demand that this simulation and the solution of a corresponding Schrödinger equation give the same Green functions for the two-meson system. For the second definition we will follow the guidelines of the finite-temperature concept [1], generalized to dynamic valence quarks.

## 2. POTENTIAL AT $T = 0$

We wish to compare the Green functions from quantum mechanics with those from field theory.

[*]Supported in part by NSF grant PHY-9409195, by FWF project P10468-PHY and by CEBAF.

In a euclidean formulation of quantum mechanics with imaginary time $\tau$ the 4-point Green function representing the two-meson system can be written as

$$G^{(4)} = \sum_n \langle y_1, y_2|n\rangle\langle n|x_1, x_2\rangle e^{-E_n\tau_f}. \qquad (1)$$

This describes the propagation of two mesons from $x_1, x_2$ at $\tau = 0$ to $y_1, y_2$ at $\tau = \tau_f$. $E_n$ are the energy eigenvalues and $|n\rangle$ the associated eigenstates. For a given potential we can solve the Schrödinger equation to obtain $E_n$ and $|n\rangle$ and calculate the sum in (1). In order to invert this process and calculate the potential from a given Green function we have to make some approximations and accept some restrictions. The first step is to set the final positions $y$ equal to the initial positions $x$. It leads to

$$G^{(4)}(d, \tau_f) = \sum_n |\langle x_1, x_2|n\rangle|^2 e^{-E_n\tau_f}, \qquad (2)$$

with a particle separation $d = x_2 - x_1$. In the next step we neglect all terms of the sum except the largest contribution, which is assumed at, say, $n = m$ for a given $d$:

$$G^{(4)}(d, \tau_f) \approx |\langle x_1, x_2|m\rangle|^2 e^{-E_m\tau_f}. \qquad (3)$$

The question emerges how $E_m$ and $\langle x_1, x_2|m\rangle$ are related to the potential $V(x_2 - x_1)$. We try to estimate the magnitude of the terms in the sum. Each term is a product of the squared modulus of the relative wave function and $e^{-E_n\tau_f}$. We can neglect the term if either the wave function or the exponential is small. Because the wave function decays exponentially "outside" the potential, it can be neglected at a given $d$ when $E_n < V(d)$. On the other hand, $e^{-E_n \tau_f}$ goes to zero if $E_n \tau_f$ becomes large and for fixed $\tau_f$ the larger $E_n$'s need not be considered. The maximum contribution for a given $d$ and $\tau_f$ can be expected near the point $E_n \approx V(d)$. In this sense the eigenvalue $E_m$ is an approximation for the potential (see also fig. 1):

$$G^{(4)}(d, \tau_f) \approx c(d) e^{-V(d)\tau_f}, \qquad (4)$$

with the positive factor $c(d) = |\langle x_1, x_2 | m \rangle|^2$. From this relation we obtain the potential as

$$V(d) \approx \frac{1}{\tau_f} \left( \ln c(d) - \ln G^{(4)}(d, \tau_f) \right). \qquad (5)$$

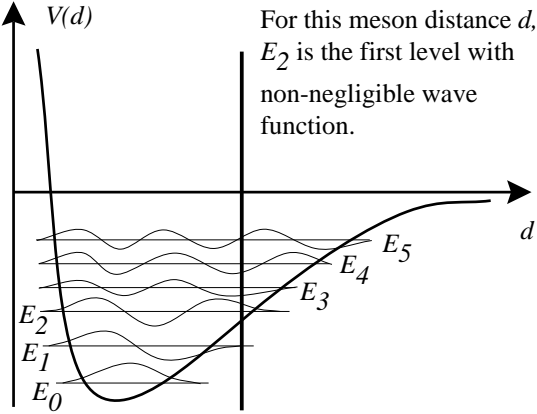

Figure 1. Estimate for the terms contributing to the sum in the spectral decomposition of the Green function (2).

## 3. POTENTIAL AT $T \neq 0$

In the case of finite temperature we take the thermodynamic free energy [1]:

$$F = -T \ln Z, \qquad (6)$$

with the temperature $T = \frac{1}{N_\tau a}$ and the partition function

$$Z = \sum_s \langle s | e^{-\frac{1}{T} H} | s \rangle, \qquad (7)$$

where $s$ contains the external sources and the vacuum states. The two-meson wave function in periodic time $\psi(x_1, x_2, \tau)$ evolves from $\psi(x_1, x_2, 0)$ according to the Green function:

$$\psi(x_1, x_2, \tau) = \int dy_1 dy_2 G^{(4)}(x_1, x_2, \tau; y_1, y_2, 0) \psi(y_1, y_2, 0). \qquad (8)$$

Proceeding as in [1], we obtain the partition function

$$Z = \langle G^{(4)}(x_1, x_2, \frac{1}{T}; x_1, x_2, 0) \rangle. \qquad (9)$$

If we assume in $F(d) = V(d) - TS(d)$ that the entropy depends only smoothly on the separation of the external sources, we can approximate the potential by the free energy.

## 4. GREEN FUNCTIONS FROM QCD

In the preceding sections we found a correspondence between the 4-point Green function for mesons and the potential in quantum mechanics. These quantum mechanical 4-point functions represent 8-point functions in QCD, because each meson consists of a quark and an antiquark. The additional colour degrees of freedom have to be saturated in a gauge invariant way. At zero temperature this is done as

$$G^{(4)\text{Meson}}(y_1, y_2, \tau; x_1, x_2, 0) = \\ G^{(8)\text{Quark}}_{aabbccdd}(y_1, y_1, y_2, y_2, \tau; x_1, x_1, x_2, x_2, 0), \qquad (10)$$

meaning that quark and antiquark are contracted at the endpoints of the 8-point function. At finite temperature the contraction is made over the periodic time:

$$G^{(4)\text{Meson}}(x_1, x_2, \frac{1}{T}; x_1, x_2, 0) = \\ G^{(8)\text{Quark}}_{abcdabcd}(x_1, x_1, x_2, x_2, \frac{1}{T}; x_1, x_1, x_2, x_2, 0). \qquad (11)$$

The 8-point functions $G^{(8)}$ can be split into 2-point functions $G^{(2)}$ (quark propagators) by fermionic integration in the path integral. This gives the same result as the application of Wick's theorem. Thus we can write [2]:

$$G^{(8)}(x_1, x_1, x_2, x_2, \tau; x_1, x_1, x_2, x_2, 0) = A + B - C - D, \quad (12)$$

with the term

$$A = G^{(2)}(x_1, \tau; x_1, 0) G^{\dagger(2)}(x_1, \tau; x_1, 0)$$
$$G^{(2)}(x_2, \tau; x_2, 0) G^{\dagger(2)}(x_2, \tau; x_2, 0) \quad (13)$$

representing the direct propagation of all quarks and antiquarks from $(x_i, 0)$ to $(x_i, \tau)$, $B$ representing the exchange of quark-antiquark pairs, and $C$ and $D$ reflecting the exchange of one quark or antiquark. The 2-point functions can be obtained from a lattice computation.

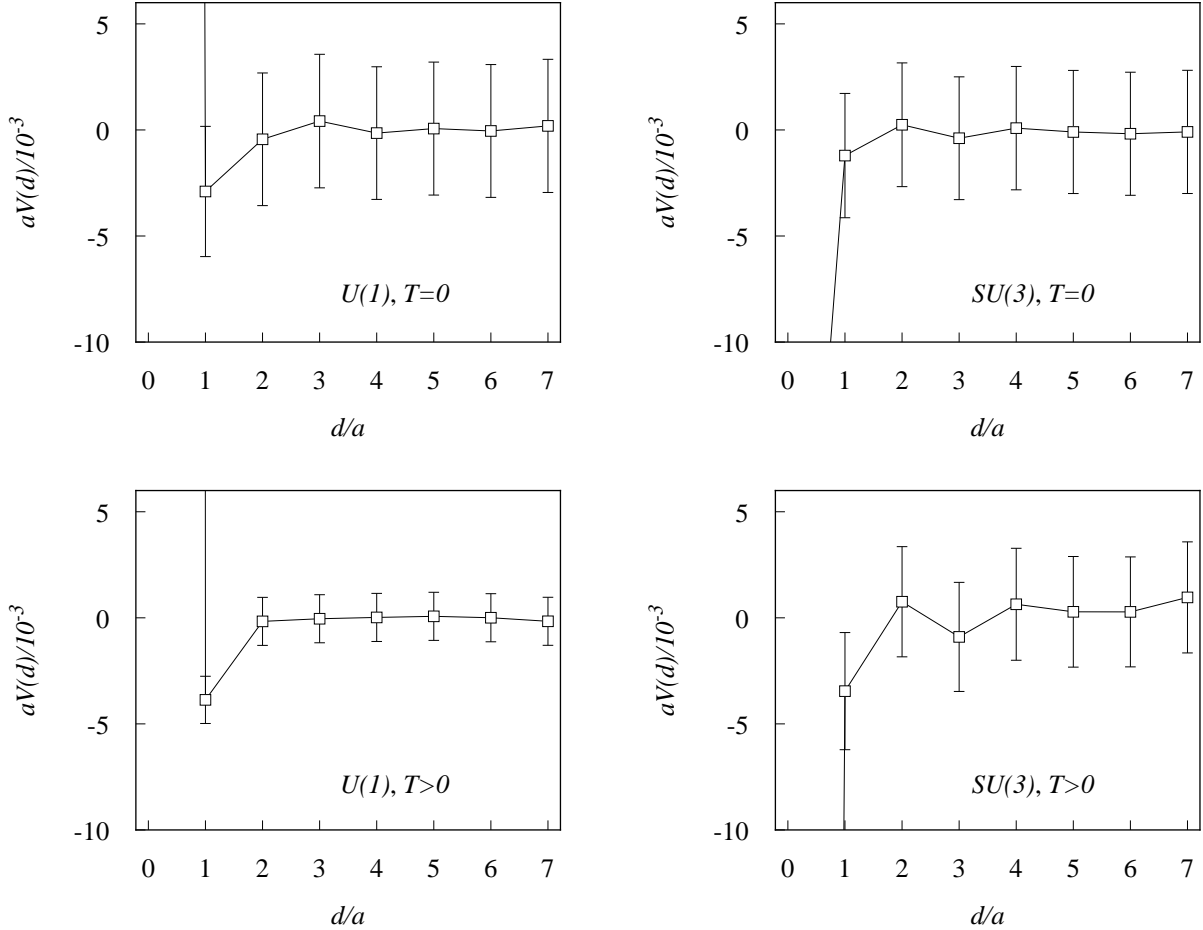

Figure 2. Meson-meson potential as a function of relative distance $d$. For $U(1)$ a short-ranged hard core is observed due to antisymmetrisation of the propagator. In the case of $SU(3)$ an attractive interaction is obtained since the nonabelian colour structure does not lead to a vanishing propagator.

## 5. RESULTS

The numerical simulation was carried out on a lattice with spatial extension $N_x \times N_y \times N_z = 8 \times 8 \times 16$ with periodic boundary conditions. The temporal extension was chosen to be $N_\tau = 16$ for zero temperature and $N_\tau = 4$ for finite temperature. The gauge fields were generated in the quenched approximation and 200 configurations separated by 50 sweeps each were analyzed. For the $SU(3)$ simulation the inverse coupling was set to $\beta = 5.6$ and for $U(1)$ to $\beta = 0.9$. The propagators were calculated for Kogut-Susskind fermions using the hopping parameter expansion. For convergence in this expansion, the quark mass was set to $m_q a = 8$.

Fig. 2 shows the results for the extracted potentials. We assume a weak $d$-dependence of $c(d) \approx$ const. The cluster value was subtracted to give $V(d \to \infty) = 0$. The potentials for $U(1)$ exhibit a hard core both for zero and finite temperature (left hand side of fig. 2) which can be understood analytically. At $d = 0$ quark exchange and direct terms cannot be distinguished and the terms $A$ through $D$ in (12) are equal. The 8-point function is thus zero and the potential infinite. For $SU(3)$ there is strong attraction at $d = 0$ (right hand side of fig. 2). This happens because the particle exchange results in a different saturation of the colour degrees of freedom and the terms $A$ through $D$ cannot cancel each other. For meson distances $d > 1$ all potentials indicate negligible interaction compared to the statistical error bars. This narrow range can be expected because of the small meson size which is inversely proportional to the quark mass chosen very large for this simulation.

## 6. HEAVY-LIGHT SYSTEM

In a second approach we inverted the fermionic matrix by a conjugate gradient algorithm using random source techniques. One of the quarks in the meson was chosen static and the light quark mass was set to $m_q a = 0.1$ [3]. An $8 \times 8 \times 8 \times 16$ lattice with 100 $SU(3)$ configurations at $\beta = 5.6$ separated by 200 sweeps was considered. The potential is extracted using (4), however the $d$-dependence of $c(d)$ is not neglected now. An exponential fit to the time behaviour of the Green function is made, with $c(d)$ and $V(d)$ as parameters. The resulting potential is drawn in fig. 3 and it turns out to be attractive again.

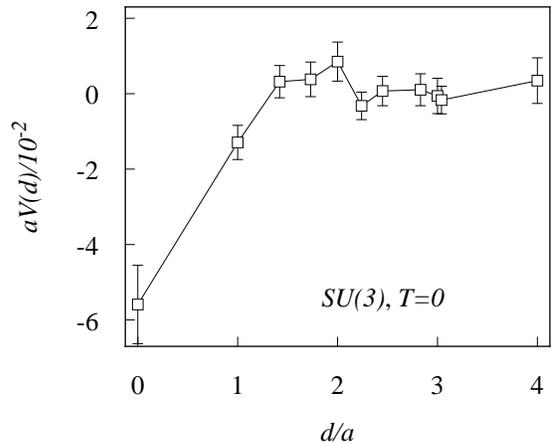

Figure 3. Meson-meson potential for heavy-light systems in $SU(3)$ at $T = 0$. Attraction is obtained as for the $SU(3)$ results in fig. 2.

## 7. CONCLUSION

We found a qualitative difference between the meson-meson potentials in compact QED and in QCD for small meson distances. In QED there is a hard core which can be understood analytically as the cancellation of the terms in a Wick expansion [2]. In QCD this cancellation is not enforced due to the sum over the colour degrees of freedom and no hard core is seen. However, all simulations were carried out in simplified models: with an unrealistically high quark mass in one case, with one quark of the meson static in the other. Our next aim is to investigate the full dynamic system with more realistic quark masses, using the conjugate gradient algorithm.